# 'Bargain your share': The role of workers' bargaining power for labor share, with reference to transition economies

**Marjan Petreski**
University American College Skopje
marjan.petreski@uacs.edu.mk

**Stefan Tanevski**
University American College Skopje
stefan.tanevski@uacs.edu.mk

## Abstract

The objective of the paper is to understand the role of workers' bargaining for the labor share in transition economies. We rely on a share-capital schedule, whereby workers' bargaining power is represented as a move off the schedule. Quantitative indicators of bargaining power are amended with own-constructed qualitative indices from textual information describing the legal enabling environment for bargaining in each country. Multiple data constraints impose reliance on a cross-sectional empirical model estimated with IV methods, whereby former unionization rates and the time since the adoption of the ILO Collective Bargaining Convention are used as exogenous instruments. The sample is composed of 23 industrial branches in 69 countries, of which 28 transition ones. In general, we find the stronger bargaining power to influence higher labor share, when the former is measured either quantitatively or qualitatively. On the contrary, higher bargaining power results in lower labor share in transition economies. This is likely a matter of delayed response to wage pushes, reconciled with the increasing role of MNCs which did not confront the workers' power rise per se, but introduced automation and changed market structure amid labor-market flexibilization, which eventually deferred bargaining power's positive effect on labor share.

1. Introduction

Trade unions play a major role in the collective bargaining process and are hence likely to influence the wage formation and labor costs. Workers' bargaining power may be consistently related with increasing labor share – the part of the value added that goes to workers, despite confounded with whether and to what extent employment shrinks after wages are the outcome of a bargaining process. Though, the primary question is if workers' power strenght, typically measured through the trade unionization rates and the collective agreements coverage rates, is an essential issue in transition economies who commenced their thorny path to market economies in the early 1990s.

Before 1990, the industrial relations systems of transition countries were characterized by central political and managerial control exercised by the state. Wages were determined according to a central pay scale based on worker's education, experience, job position and so on, but collective bargaining and unions were instruments in the hands of the party machine rather that a real workers' power (Jackman, 1994; Borisov and Clarke, 2006). With the collapse of the central-planning economy, efforts were made to develop industrial relations typical for a market economy, and all transition countries – with a various degree, have started to move away from a centralized wage setting system, towards a collective bargaining system in the enterprise sector (Cazes, 2002).

While wages were controlled, workers in the socialist systems enjoyed a fairly high degree of employment protection in their jobs, despite freely able to move across jobs and employers. Though, the high degree of employment protection, combined with high wage compression, led to extreme labor rigidity and inefficient labor allocation. Many of the transition economies embarked on labor-market flexibilization over the 1990s, as part of the plans for structural adjustment to facilitate economically competitive enterprises which yet guaranteed employment protection for workers comparable with that in advanced economies (Lehmann and Muravyev, 2011). Indeed, new labor codes at the onset of the transition period provided the rights to association and to strike, allowed the main elements of the employment contact, including wages, to be determined by a collective bargaining between workers represented through the trade unions and employers at the level of enterprise. This generally set the basic legal framework in accordance with the practice in most of the market economies, while leading to a substantial moderation of workers' protection, in general, which was made possible also due to the weakening of trade union power.

Limited available data indeed suggest that transition economies presently have fairly flexible labor market in place. The OECD Strictness of Employment Protection Index reveals numbers as low as 1.67 in Kosovo and 1.7 in Serbia, measured on a 0-6 scale whereby the higher the number the stricter the employment protection. Then, this is followed by 2.14 in North Macedonia and Albania, and 2.52 in Bosnia and Herzegovina, which are very similar to Slovenia, 2.08 and Croatia, 2.48, which are already members of the EU, hence already advanced post-transition economies. Yet, countries like the fast transitioning Czechia and Latvia have been more rigid in terms of employment legislation, an index of 3.26 and 3.02, respectively, and very similar to slow transitioning Russian Federation and Kazakhstan, 3.06 and 3.32, respectively. Low employment protection may reflect that workers' power in transition economies has been on the low end.

Trade union density has been on the decline likewise. Actually, some transition economies exert very low rates of such density, e.g. 6% in Estonia and 7.4% in Lithuania, which actually dropped from 7.6% and 10% a decade ago, respectively. However, drops were much deeper in the other transition economies with larger densities, for example from 49.5% to 36.9% in Albania and from 38.4% to 20.6% in Armenia, both over the last decade. Lehmann and Muravyev (2011) consider such trends since the early 1990s to be a manifestation of the reshaping of the collective bargaining systems of transition economies, to resemble those of the advanced market economies. Still, such declining trade union density rates may not reflect neither the fact that collective agreement coverage rates may be rather high nor that collective bargaining legislation has been evolving to ensure a level playing field favoring true dialogue that is beneficial for workers' welfare.

We ponder of the workers' welfare as the share of the value added that goes in the hands of workers, widely known as the 'labor share'. Surprisingly, though, the labor share has been on the rise in wide majority of transition economies, opposing the early Kaldor's (1961) constancy of labor share as a "stylized fact of growth" or the more recent declining labor share widely documented (Acemoglu, 2003; Atkeson, 2020). Rather, increases have been ranging from very large to moderate; for example, the labor share was 27.5% in 1990 in Bosnia and Herzegovina and rose to 55.6% in 2020. In many transition economies, the three-decennial increase has been as much as 20 percentage points, though ranging to modest in countries like Hungary whereby it moved from 39.8% in 1990 to 43.8% in 2020. Various causes may have contributed to the behavior of the labor share: how labor and capital were combined during the process of transition; policies pursued by governments (e.g. of wage moderation interfering soaring inflation); global factors and the influx of multinational corporations; automation and new technologies; product market structures; workers' power and others. A comprehensive review of the literature on what potentially explained labor share movements is provided by Grossman and Obstfeld (2022).

We attempt to synthetize the argumentation around the latest factor: workers' power. Bantolila and Saint Paul (2003) posit that whenever unions force wages above the competitive rates and firms are left free to decide on employment, the labor share will be higher with unions than without them when the elasticity of substitution between capital and labor is less than unity. Cauvel and Pacitti (2022), moreover, argue that labor bargaining power is actually the common denominator of all determinants of the labor share: the business cycle (structural economic factors) affects the labor share because labor bargaining power is pro-cyclical, the influx of multinational corporations in small economies (globalization factors) affects the labor share through exhausting otherwise weakly unionized labor through long hours and minimum wages (Petreski, 2020), while the adoption of automation and capital-based technologies deteriorates worker's power and lowers their revenue share, inter alia through inducing deunionization (Holmes et al. 2012). Stansbury and Summers (2020), Cárdenas and Fernández (2020), Ramskogler (2021), Brancaccio et al. (2018) and Ciminelli et al. (2022) argue that institutional changes – deunionization, collective agreements, strike activity, labor law protection strength, labor standards and law enforcement - are the main cause of lower worker bargaining power and thus of labor's share of income.

The empirical literature examining the role of workers' bargaining power for the labor share has been rather focused on advanced economies, yet without strong undivided conclusions. Bantolila and Saint Paul (2003), for 12 OECD countries, used the number of labor conflicts as a proxy for the collective bargaining power and found that they reduce the labor share, possibly suggesting delayed response to wage pushes, despite the finding was marginally significant. Farber et al. (2021), Cauvel and Pacitti (2022) and Stansbury and Summers (2020), on the other hand, documented a positive relationship between workers' bargaining power – though captured through various proxies - and labor share in the US, despite such finding was frequently not corroborated, e.g. Elsby et al. (2013) found rather small explanatory power of the unionization rate for the workers' share in income. To our knowledge, only Petreski (2021) puts labor share in the context of market bargaining power in transition economies, captured through the skill composition of the idle labor, but finds a significant positive relationship only in high-skill industries of the advanced transitioners.

The objective of the paper is to understand the role of workers' bargaining for the labor share in transition economies. Given scarce standard and quantitative indicators of the strength of workers' bargaining, we make an expanded use of self-constructed indicators that capture also the 'qualitative side' of workers' strenght in the bargaining process, mainly through indexing the provisions in the relevant labor laws; and do this in a comparative perspective. As such, the paper brings a couple of novelties to the current sparse of knowledge. *First*, the paper puts the issue in a theoretical model whereby the economy is allowed to deviate from a standard schedule depicting the labor-capital relation due to existence of bargaining between workers and employers. While our contribution is mainly applicative, we consider the theoretical framework used in such a context for the first time. *Second*, little is known about the labor share in transition economies; part of the recent literature is focused on the role of globalization factors (e.g. Petreski, 2021), while the area of collective bargaining power of workers in its broad sense (beyond unionization: representation, facets of and clauses in collective agreements and so on), including the aspects which are not measured in numbers per se, has not triggered much empirical attention to our knowledge. This is important given two currents shaped bargaining: the old one inherited from how the system worked under socialist economy, and the new one determined by the exposure of these small and open economies to the global flows of capital, and by labor and skill scarcity acquaint with the massive emigration in the last decade or two. Moreover, given the time-limitation of our data, we rather focus on the comparative perspective of our transition-economies results versus those of the other (mostly) advanced economies. *Third*, the paper is the first to make extensive use of the legislative setup in the area of workers' bargaining power as per the provision of the relevant laws, hence rather capturing the quality of the process, notwithstanding the fact that this prevented that we analyze the matter in a time perspective. *Fourth*, given the data scarcity on the matter, we consider a contribution the maximum unfolding of the data and variables, also through the attempt to pursue a type of endogeneity-correction estimation.

The rest of the paper is organized as follows. Section 2 displays the theoretical framework we use for our empirical examination. Section 3 presents the data used, with particular emphasis on the own-created indices that capture various qualitative aspects of the workers' bargaining power. Section 4 debates the underlying methodological issues. Section 5 presents the results and offers a discussion. Section 6 concludes.

2. **Theoretical underpinnings**

To arrive at a theoretically-sound empirical model about the labor share, we are guided by the theoretical framework of Bantolila and Saint Paul (2003), which nests bargaining within a model that explains the relationship between the labor and capital. We, however, refrain from showcasing the mathematical derivation, since our objective in this paper is rather empirical. The model assumes a departure from a standard Cobb-Douglas production function, as the latter adopts a one-to-one relation between the labor share and the capital-output ratio. However, even if the production function is not of a Cobb-Douglas type, Bantolila and Saint Paul (2003) show that there is a stable relation between the labor share and an observable variable, the capital-output ratio, which they refer to as the share-capital (SK) schedule or curve, shown on Figure 1.

### Figure 1 – The SK schedule

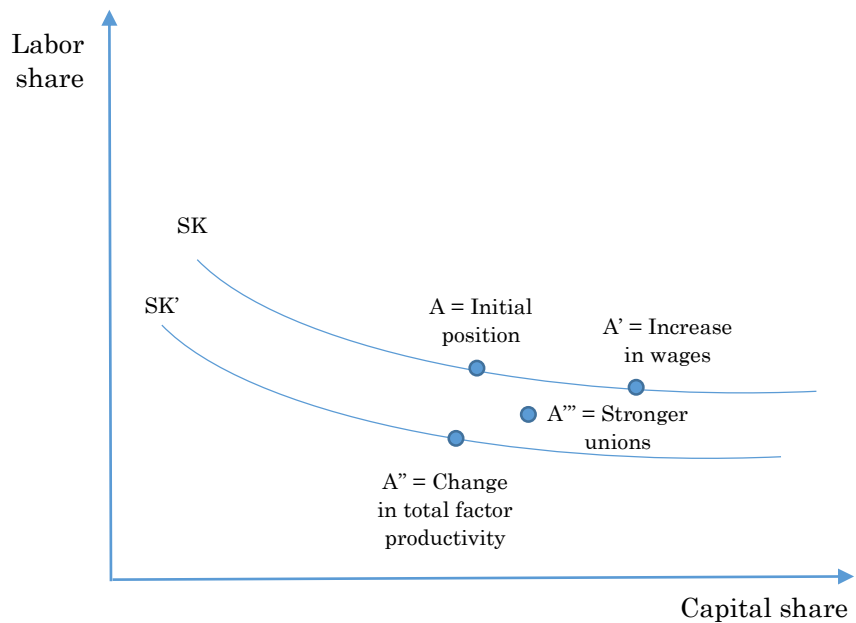

*Source: Adapted from Bantolila and Saint Paul (2003).*

The economy is initially positioned in point A on the SK curve, a point where the marginal product of labor is equal to the real wage. Any change in the prices of factors on the graph's axes – i.e. wages or the interest rates – or quantities, or in labor-augmenting technical progress, causes a shift along the SK schedule, to A'. However, changes in the labor share may be explained by such changes in the capital-output ratio or may remain unexplained – variations captured by the residual, i.e. due to other factors. The response of the labor share to the capital-output ratio is related to the elasticity of substitution between labor and capital: their strong complementarity means that an increase in the capital-output ratio is associated with a larger labor share, and vice versa.

However, the technical progress may be capital-augmenting as well, in which case the relation between the labor and capital is no longer stable. Under capital-augmenting technical progress, productivity shifts (if not constant) move the SK schedule to SK' on Figure 1 and the economy will position at A". The upward or the downward movement of the SK schedule will though depend on the characteristics of the production function.

Furthermore, there are factors which create a wedge between the real wage and the marginal product of labor. They do not necessarily affect the relation between the employment elasticity of output and capital-output ratio, but create a gap between the former and the labor share. Hence, such factors do not shift the SK curve, but put the economy off that schedule, to point A''' on Figure 1. One source of such gap is the bargaining, on which we focus. Labor share increases are routinely interpreted as increases in the bargaining power of workers, which is understood as a necessity of the employment to decline. But the relation depends on the bargaining model. If workers and employers bargain only over the wages, while employment is a residual (right-to-manage model), then changes in the bargaining power will move the labor share along the SK curve, in a direction that depends on the elasticity of substitution between labor and capital. If workers and employers bargain around both wages and employment (efficient bargaining model), then an increase in bargaining power of workers will affect the labor share in a short run, but not employment, while over the long run, adjustment of the capital stock would imply that it also affected employment. The shift will be off the curve, i.e. above it, reflecting that workers are paid more than their marginal product. In such cases, the increases in worker's bargaining power reduce the sensitivity of the labor share to the capital-output ratio.

Bantolila and Saint Paul (2003) present a mathematical derivation for two other sources of such gap: the markups reflecting product market power and the labor adjustment costs. If markups are constant, then the relation between the labor share and capital is stable, but their counter-cyclicality, for example, will imply that the labor share is pro-cyclical given the capital endowment. Labor adjustment costs – e.g. needs for training of workforce in case of hiring or severance payments in case of firing – push the marginal cost of labor above or below the wage, despite the expected future marginal adjustment costs may depend on the degree of uncertainty in the economy.

Our empirical specification, therefore, is as follows:

$$lnS_{L,ijt} = \beta_0 + \beta_1 lnk_{ijt} + \beta_2 lnTFP_{ijt} + \beta_3 \Delta lnn_{ijt} + \sum_{k=1}^{n} \gamma_k cb_{jt,k} + v_{ijt} \qquad (1)$$

whereby $lnS_{L,ijt}$ is the logarithm of the labor share in value added in industry $i$ in country $j$ in time $t$. $lnk_{ijt}$ is the logarithm of the capital-output ratio; $lnTFP_{ijt}$ is the logarithm of the total factor productivity, aimed to capture the capital-augmenting technical progress, while the other terms capture the deviations of the wage from the marginal product of labor. $\Delta lnn_{ijt}$ is the industry net growth rate of the number of employees, while $cb_{jt,k}$ stands for a vector of variables representing workers' bargaining power, which vary only by country. $\Delta lnn_{ijt}$ is not very well measured from the viewpoint of the theory, while pure markups are difficult to be measured. These two are anyway not under the spotlight of this analysis and hence we include them to the extent possible to alleviate any omitted-variable bias. $v_{ijt}$ is the residual term which is assumed to have the standard properties.

Therefore, our model (1) focuses on three sources of variation in the labor share: movements along the SK schedule, i.e. shocks whose effect on the labor share is entirely mediated by the capital-output ratio, such as changes in interest rates, wages or the labor-augmenting technical progress; one shifter of the SK curve, the capital-augmenting technical progress; and two sources of movements off the SK schedule, namely changes in labor adjustment costs and in workers' bargaining power, of which we are mainly interested in the latter.

### 3. Data

We use data from the UNIDO Industrial Database, namely: the labor share, capital-output ratio, number of employees, for manufacturing industrial branches at the two-digit ISIC classification. The labor share is constructed by dividing the amount paid as wages and wage supplements to workers with the value added. The capital share is the ratio of the gross fixed capital formation per industrial branch and the total output. The total factor productivity is a derived variable in a standard growth accounting exercise in which the value added, number of workers and fixed capital formation per industry feature as inputs.

We make use of the entire UNIDO Industrial Database, namely for the years between 2000 and 2021, for all 23 branches of manufacturing industry and for 162 countries. Still, the panel is strongly unbalanced, since many observations are missing, either at the level of industrial branch, of particular years, or of particular countries. Actually, the share of missing data for the value added and the output, two key variables we use, is severe, as documented in Table A1 in the Annex.

The issue becomes even starker when we attempted the usage of the bargaining indicators. For this study, we make use of two strands of indicators. The first is the ones of the ILO Industrial Relations Database, which has on disposal six indicators: Trade union density rate, Collective bargaining coverage rate; Number of strikes and lockouts; Days not worked due to strikes and lockouts; Workers involved in strikes and lockouts in total and for manufacturing only. These indicators are available for the time span 2009-2020, for a set of countries whose number ranges between 56 and 123 for the various indicators, with many gaps, hence undermining our ambition to construct a truly global panel. Table A2 in the Annex provides an overview of this set of variables.

The second strand of indicators comes from the ILO Legal Database on Industrial Relations (IRLex), which provides accessible summaries and legal texts on industrial relations across ILO member states. Hence, this database is rather textual/qualitative overviews of how collective bargaining is regulated in each country, representing the current situation, i.e. without any time component. It describes the legal aspects of seven domains, namely: Regulatory framework, Organizations and their administration, Legislative protection of workers' and employers' organizations, their members and representatives, Information and consultation at the workplace, Labor disputes and their resolution, Tripartite social dialogue, and Collective bargaining. In this paper, we focus on the latter two. We collect qualitative information for the 70 available countries, which entirely match the ultimately-usable data of the countries of our quantitative indicators.

Our methodology consists of coding the qualitative information into ordered categorical variables within the two thematic domains, in the manner that the context in each country is labelled from representing a 'very unfavorable situation for collective bargaining' to 'very favorable situation for collective bargaining', on a scare from 1 to a maximum of 6. Wherever 6 gradients were not possible, variables were rescaled so that ultimately each aspect of collective bargaining is evaluated within the 1-6 span, hence abandoning the ordered character of the variables so as to be able to ultimately average up in an index. This way, we created five indices, as follows:

- *Basic requirements in collective bargaining* covering representation requirements for trade unions to negotiate a collective agreement at national level, requirements for bargaining in good faith, and registration of collective agreements;

- *Basic characteristics of collective agreements* covering initiating the agreements, duration, their amending and termination;

- *Non-signatory parties and exemptions* covering automatic/non-automatic extension to non-signatory parties, the resting power to extend, and the mechanisms for exemption from binding effects of an extended collective agreement;

- *Institutional/legal characteristics of tripartite dialogue* covering the number of institutions acting as an umbrella for tripartite dialogue, the legal status of the institution, the founding instrument, and the legal effect of the institutions;

- *Members/representatives in tripartite dialogue* covering the number of members in the founding body, and the representatives other than workers, employers and government included in the institution.

All indicators within an index obtained equal weighting. Table A3 in the Annex provides an overview of the coding related to this set of variables, while Table A4 presents descriptive statistics.

Given: i) gaps in the ILO Industrial Relations Database; and ii) current state of art derived from the ILO IRLex database, i.e. impossibility to track changes over time in the qualitative facets of bargaining, we reduced our sample to a cross-section one. We took the current value of each variable, mainly corresponding to the year of 2020, but for the cases when this was not available, we took the most recent value going back for a maximum of five years. For the variables which offered more than the current value, we also constructed the previous value, i.e. the one preceding the value we took as the current; in most cases, this was the previous year, but wherever this was not available we took the most recent one going back for a maximum of five years.

### 4. Methodology

The usual methods for estimating our equation (1) involve instrumental-variable estimators, including Arellano-Bover-type of, which take into account the potential endogeneity in the regression, as well as introduce some dynamics. This is our starting point, yet not the one that is used for the core objective of the study. Namely, we estimate equation (1) by relying on FE panel estimator, IV/2SLS estimator and Arellano-Bover system-GMM estimator, whereby individual units of observation are industry-country pairs, yet with the variables that capture movements along the SK schedule, through using the capital-output ratio, and a shift of the SK curve, through using the TFP index. For this initial estimate, we abstract from the movements off the SK schedule, because of the data deficiencies spelled out in the previous section. Namely, our key interest is in the latter, but by first introducing panel-based estimates, we aim to calculate the coefficients on the capital-output ratio and the TFP index, which will serve as yardsticks once we reduce our calculation to the cross-section level.

Once our dataset reduces at the cross-sectional level, we do not ignore endogeneity either, the source being unobservables which imply correlations between the labor share and the shocks onto it. For the variables for which time dimension was available, we retain the value of the previous period to be used as instrument, hence building on the notions of the system-GMM panel estimator, as well enrich our instruments' portfolio through constructing a variable measuring the number of years for which countries have had the ILO Collective Bargaining Convention, 1981 (No. 154) in force since ratification. It is not uncommon in the literature to rely on such type of sources of exogenous variation; for example, Farber et al. (2018) for the US used the passage of the Wagner Act in 1935, which legalized union organization, and the establishment of the National War Labor Board in 1942, which promoted unionization in establishments receiving defense contracts during World War II, as plausible exogenous sources of variation in states' unionization rates.

Yet, one must be clear of what cross-sectional specification cannot achieve. Nakamura and Steinsson (2018) argue that such estimates of macroeconomic effects neglect general-equilibrium adjustments that affect all units of interest in a similar fashion. In our context, this would mean that the constant in the cross-sectional regression of factor shares absorbs the effect of relative factor prices on input choices. In other words, economy-wide variabilities as the average wage, the interest rate, the amount of human capital etc. evolve endogenously in a dynamic equilibrium, preventing that a historical episode is analyzed. Yet, a cross-sectional specification saves for the problem of identification due to pervasive simultaneity present in time-series and panel specifications.

So, in both panel and cross-sectional specifications, we treat the right-hand side variables as potentially exogenous and characterize endogeneity in the panel regressions, as usual through the composition of the error ($v_{ijt}$ in eq. 1) by the correlation of the industry-country fixed effect with the explanatory variables and by the individual shock (which is period-specific when time component is included), which captures measurement errors and unobservables as e.g. markups. We use as instruments the lagged values of the potentially endogenous variables for whom time dimension was available and the variables capturing the time since the ratification of the relevant ILO conventions. We use a linear IV specification of the 2SLS type, as well as Arellano and Bover's (1995) system estimator. We rely on the standard set of tests for the overidentifying restrictions to validate instruments. We also report a statistic for the absence of second-order serial correlation in the first-differenced residuals for the Arellano-Bover estimates.

## 5. Results
### 5.1. The basic empirical model with time component

We first provide estimates of our basic model (1) by dwelling on two critical components: the movements along the SK curve and the shifts of the SK curve. The results are presented in Table 1. IV-based specifications are fine according to the respective tests at the bottom of the table, while the significances onto the lagged dependent variable do not provide a clear case for the usage of a dynamic specification. The coefficient for the capital-output ratio implies a capital-labor elasticity in the range between 1.13 and 1.19 (taking the average elasticity of the labor demand with respect to the wage in Hamermesh, 1993 of -0.39). For the transition economies, the range is slightly higher, between 1.16 and 1.23, and the coefficient on the capital-output ratio is insignificant in the IV-based specification. The estimated range is statistically different from the Cobb-Douglas value of 1. The coefficient on the total factor productivity, meant to capture the effect of the capital-augmenting technical progress on the labor share, is negative, despite the magnitude is difficult to interpret due to variable defined through a residual in a Cobb-Douglas specification. As its sign is the same as the sign of the capital-output ratio, the TFP is argued to be strictly capital-augmenting, including in transition economies only.

Table 1 – Results of the basic model with time component

|  | Dependent variable: Log labor share | | | | | |
|---|---|---|---|---|---|---|
|  | All economies | | | Transition economies | | |
|  | FE | IV | Arellano-Bond | FE | IV | Arellano-Bond |
|  | (1) | (2) | (3) | (4) | (5) | (6) |
| Capital-output ratio (log) | -0.383*** | -0.323*** | -0.482*** | -0.412*** | -0.213 | -0.577*** |
|  | (0.016) | (0.093) | (0.032) | (0.033) | (0.283) | (0.061) |
| TFP (log) | -1.030*** | -1.109*** | -1.164*** | -0.965*** | -0.669** | -1.182*** |
|  | (0.059) | (0.098) | (0.065) | (0.064) | (0.262) | (0.121) |
| Lagged labor share (log) |  |  | 0.134 |  |  | 0.200*** |
|  |  |  | (0.088) |  |  | (0.070) |
| Constant | -2.853*** |  |  | -2.915*** |  |  |
|  | (0.086) |  |  | (0.159) |  |  |
| Observations | 2,715 | 1,801 | 1,932 | 1,204 | 855 | 930 |
| R-squared | 0.687 | 0.690 |  | 0.571 | 0.551 |  |
| Number of fixed effects | 403 | 292 | 324 | 152 | 129 | 146 |
| Hansen test of overid. restrictions (p-value) |  | 0.595 | 0.387 |  | 0.454 | 0.215 |
| Arellano-Bond test for AR(2) |  |  | 0.370 |  |  | 0.098 |

Source: Authors' calculations.
Note: Country-industry pairs and year dummies used as fixed effects. Past values of the right-hand side variables used as exogenous instruments. Standard errors are provided in parentheses and are robust to arbitrary heteroskedasticity and serial correlation.

### 5.2. Labor share under bargaining measured quantitatively

We turn to the results from the cross-sectional regression, in this section, by first introducing the quantitative bargaining indicators, in **Table 2**. Note that we instrument all the regressions with the most recent past value of the log trade union density rate (the quantitative indicator with the most observations) and the indicator about the time since the C154 was ratified (whereby zero stands for this convention not being ratified yet). The Hansen test for the instruments validity is presented towards the bottom of the table and suggests that instruments are valid.

Results based on both FE and IV estimators offer similar and consistent conclusions. First, the capital-output ratio coefficient implies a capital-labor elasticity in the range between 1.08 and 1.23, so in a slightly expanded interval than in **Table 1**, but fully consistent with the findings based on a panel construct. Likewise, the TFP is significant and consistently negative, corroborating its strictly capital-augmenting role. Changes in the employment rate are significant only in the specifications based on OLS-FE and the coefficient is negative, suggesting that a growing employment reduces the labor share, which is expected given it captures a labor adjustment cost. However, the significance is lost when endogeneity-corrected estimates are presented.

Turning to the coefficients capturing the bargaining in quantitative sense, we conclude that almost all of them are statistically significant, positive and of similar magnitudes, both within the OLS-FE and IV specifications. This suggests that bargaining power of workers has a significant role to play for the labor share; namely, the stronger the bargaining, the higher the labor share. The coefficients could be interpreted in the following fashion: an increase of the trade union density rate by 1% (given we used logs for the rate), increases the labor share by 0.09%, which is not a small change. Or, increasing the collective bargaining coverage rate by 1%, increases the labor share by 0.08%. And so on. In the IV-based estimates, these coefficients become larger (despite the one on the trade union density is insignificant). Overall, they suggest that whenever the bargaining power is increased by 10%, the labor share increases by, on average, 1.2% to 1.5%.

**Table 3** presents the same results for the transition economies group; however, due to the rather small sub-sample of transition economies, we only introduce a transition-country dummy and its product with the bargaining indicators for three of them: trade union density rate, collective bargaining coverage rate and number of strikes and lockouts. Previous conclusions related to the capital-output ratio, TFP and the employment growth are corroborated.

The transition-country dummy is significant in some of the specifications and positively signed, but with volatile magnitude. With some caution, it may still be used to reconcile the known fact that transition economies experienced increasing labor share compared to the non-transition economies, whereby the labor share has been mostly on the decline by the time spot of our cross-section data. Then, the basic bargaining indicators corroborate the finding from the full sample in **Table 2**: higher bargaining power in non-transition economies, majority of which advanced ones, is associated with higher labor shares. On the other hand, the cross-product of the bargaining indicator and the transition-country dummy is significant in three out of six cases, in all being negative, and the coefficient is larger than the basic bargaining variable. Hence, even if they are observed together, the sum of the coefficients remains negative. It suggests that in transition economies, the bargaining power has been either related with smaller labor shares or it did not matter.

While this result is affected by the sample size, given all limitations we faced with the data availability, to our opinion, it rests with a general importance for transition economies. One plausible explanation of the negatively signed bargaining power of workers for their share in income rests with the notion of delayed response of the labor share to wage pushes attained through workers' concerted effort, theoretically elaborated in Caballero and Hammour (1998). Another less plausible explanation would be that workers and employers bargain only over the wages, and then set employment, which results in declining labor share when taken in aggregate. However, this would show up as a move along the SK curve, and empirically the coefficient would be rather insignificant than negative (which is the case in half of the estimated coefficients on transition economies, but more likely due to sample-size issues). Practically observed, the finding neatly describes the reality: transition economies have been exposed to the influx of the foreign capital after the dissolution of the former socialist system, which did not confront the workers' power rise, hence not directly impacting the bargaining power reflected in unionization rates, but operated through its underlying channels, like the rapid automation multinationals introduced in their settlements in the host countries, and the change in market structure towards larger and more powerful firms, ultimately overturning the influence of bargaining power onto labor share or adjourning its potentially positive response.

Table 2 – Results of the full model with quantitative bargaining indicators, all countries

| | Dependent variable: Log labor share | | | | | | | | | | | | | |
|---|---|---|---|---|---|---|---|---|---|---|---|---|---|---|
| | Fixed effects estimator | | | | | | | IV/2SLS estimator | | | | | | |
| | (1) | (2) | (3) | (4) | (5) | (6) | (7) | (8) | (9) | (10) | (11) | (12) | (13) | (14) |
| Capital-output ratio (log) | -0.176*** | -0.214*** | -0.440*** | -0.0673 | -0.0818 | -0.0764 | -0.493*** | -0.186 | -0.109 | -0.597*** | -0.209*** | -0.163 | -0.214** | -0.598*** |
| | (0.044) | (0.081) | (0.043) | (0.105) | (0.117) | (0.110) | (0.060) | (0.146) | (0.281) | (0.177) | (0.080) | (0.125) | (0.087) | (0.079) |
| TFP (log) | -0.577*** | -0.676*** | -0.927*** | -0.426*** | -0.458*** | -0.500*** | -0.813*** | -0.406 | 0.0846 | -1.174*** | -0.819*** | -0.725*** | -0.832*** | -0.748*** |
| | (0.063) | (0.119) | (0.079) | (0.162) | (0.151) | (0.150) | (0.105) | (0.261) | (0.703) | (0.237) | (0.139) | (0.191) | (0.116) | (0.158) |
| Employment growth (log) | -1.458*** | -2.167* | -1.285** | -2.369* | -1.904* | -1.970** | -1.663*** | -13.86 | -25.34* | 3.451 | -2.473 | -0.808 | -2.567 | 3.579 |
| | (0.546) | (1.139) | (0.513) | (1.305) | (1.082) | (0.916) | (0.413) | (5.930) | (14.590) | (7.183) | (1.690) | (4.702) | (2.435) | (2.542) |
| Trade union density (log) | | 0.0863* | | | | | | | -0.372 | | | | | |
| | | (0.045) | | | | | | | (0.439) | | | | | |
| Collective bargaining cov. rate (log) | | | 0.0833*** | | | | | | | 0.141*** | | | | |
| | | | (0.020) | | | | | | | (0.049) | | | | |
| Number of strikes and lockouts (log) | | | | 0.0522*** | | | | | | | 0.131*** | | | |
| | | | | (0.016) | | | | | | | (0.017) | | | |
| Number of days not worked due to strikes (log) | | | | | 0.014 | | | | | | | 0.145** | | |
| | | | | | (0.013) | | | | | | | (0.069) | | |
| Number of workers inv. in strikes, total (log) | | | | | | 0.0521*** | | | | | | | 0.0694*** | |
| | | | | | | (0.012) | | | | | | | (0.009) | |
| Number of workers inv. in strikes, manuf. (log) | | | | | | | 0.0981*** | | | | | | | 0.122*** |
| | | | | | | | (0.009) | | | | | | | (0.020) |
| Constant | -1.805*** | -2.240*** | -2.801*** | -1.655*** | -1.724*** | -1.804*** | -3.471*** | -2.297*** | -1.364 | -3.225*** | -2.453*** | -2.793*** | -2.366*** | -3.781*** |
| | (0.129) | (0.275) | (0.134) | (0.326) | (0.325) | (0.330) | (0.231) | (0.419) | (1.188) | (0.365) | (0.242) | (0.530) | (0.242) | (0.342) |
| Observations | 365 | 236 | 176 | 160 | 133 | 135 | 78 | 207 | 207 | 167 | 137 | 118 | 118 | 78 |
| R-squared | 0.437 | 0.519 | 0.648 | 0.418 | 0.391 | 0.465 | 0.726 | -2.024 | -9.4 | 0.329 | 0.39 | -0.676 | 0.471 | 0.111 |
| Hansen test (p-value) | | | | | | | | 0.256 | 0.787 | 0.226 | 0.862 | 0.0711 | 0.725 | 0.556 |

Source: Authors' estimates. The trade union density rate from the most recent past period and the log number of days since C154 has been ratified (0 if not ratified yet) used as exogenous instruments. Standard errors are provided in parentheses and are robust to arbitrary heteroskedasticity.

Table 3 – Results of the full model with quantitative bargaining indicators, transition countries

| | Dependent variable: Log labor share | | | | | | | |
|---|---|---|---|---|---|---|---|---|
| | Fixed effects estimator | | | | | IV/2SLS estimator | | |
| | (1) | (2) | (3) | (4) | (5) | (6) | (7) | (8) |
| Capital-output ratio (log) | -0.175*** | -0.191** | -0.435*** | -0.0683 | -0.290*** | -0.498 | -0.539*** | -0.222** |
| | (0.044) | (0.091) | (0.041) | (0.111) | (0.072) | (0.303) | (0.091) | (0.090) |
| TFP (log) | -0.583*** | -0.619*** | -0.878*** | -0.425*** | -0.730*** | -0.601* | -0.960*** | -0.840*** |
| | (0.066) | (0.121) | (0.083) | (0.162) | (0.119) | (0.361) | (0.138) | (0.128) |
| Employment growth (log) | -1.503*** | -1.754 | -1.126** | -2.367* | -5.855*** | -19.55** | 3.004 | -2.004 |
| | (0.579) | (1.254) | (0.517) | (1.339) | (1.427) | (9.655) | (2.056) | (2.595) |
| Transition economies (dummy) | -0.0674 | 2.004*** | 0.657*** | . | 0.214** | -5.599 | 2.088*** | . |
| | (0.081) | (0.673) | (0.240) | . | (0.095) | (4.643) | (0.740) | . |
| Trade union density (log) | | 0.137* | | | | -0.98 | | |
| | | (0.073) | | | | (0.629) | | |
| *Transition economies* | | -0.595*** | | | | 2.057 | | |
| | | (0.203) | | | | (1.607) | | |
| Collective bargaining cov. rate (log) | | | 0.0964*** | | | | 0.252*** | |
| | | | (0.024) | | | | (0.061) | |
| *Transition economies* | | | -0.187*** | | | | -0.684*** | |
| | | | (0.069) | | | | (0.213) | |
| Number of strikes and lockouts (log) | | | | 0.0524*** | | | | 0.135*** |
| | | | | (0.015) | | | | (0.019) |
| *Transition economies* | | | | 0.00878 | | | | 0.0544 |
| | | | | (0.119) | | | | (0.122) |
| Constant | -1.778*** | -2.318*** | -2.801*** | -1.659*** | -2.454*** | -1.124 | -3.153*** | -2.500*** |
| | (0.140) | (0.219) | (0.131) | (0.349) | (0.201) | (0.900) | (0.238) | (0.258) |
| Observations | 365 | 236 | 176 | 160 | 207 | 207 | 167 | 137 |
| R-squared | 0.439 | 0.56 | 0.66 | 0.418 | 0.257 | -4.807 | 0.365 | 0.385 |
| Hansen test (p-value) | | | | | . | 0.551 | 0.94 | 0.79 |

*Source: Authors' estimates. The trade union density rate from the most recent past period and the log number of days since C154 has been ratified (0 if not ratified yet) used as exogenous instruments. Standard errors are provided in parentheses and are robust to arbitrary heteroskedasticity.*

### 5.3. Labor share under bargaining measured qualitatively

In this section, we refer to the results from the qualitative indicators of the collective bargaining, in **Table 4**. The capital-output ratio coefficient implies a capital-labor elasticity in the range between 1.07 and 1.25, very consistent with the findings based on the quantitative indicators (**Table 2**). The conclusions related to TFP and changes in the employment are further fully corroborated.

Almost all coefficients capturing the bargaining in qualitative sense are statistically significant and positive, with a higher magnitude in the IV specifications. This suggests that bargaining power of workers, captured through qualitative indicators, has a significant role to play for the labor share; namely, the stronger the bargaining as per its legal prescriptions, the higher the labor share. The coefficients could be interpreted in the following fashion: an increase of the power of workers in the social dialogue related to e.g. collective agreements duration, amendments and termination by one tenth of a unit (on a 1-6 scale), increases the labor share by 3.7%. Or, increasing the collective bargaining power through enriching founding members of unions and securing representation of other stakeholders by one tenth of a unit, increases the labor share by 2.8%.

**Table 5** presents the same results for the transition economies group. The basic bargaining indicators corroborate the finding from the full sample in **Table 4**: higher bargaining power is associated with higher labor shares. Conversely, the cross-product of the qualitative bargaining indicator and the transition-country dummy is negative, and the coefficient is larger than the basic bargaining variable. It reveals that in transition economies, the bargaining power has been either related with smaller labor shares or it did not matter, a conclusion that we also reached through the quantitative indicators and we fully verify it here.

On top of the discussion in Section 5.2 about how the influx of multinational corporations potentially shaped the relation between bargaining and labor share in transition economies, this finding is further significant from the viewpoint of the labor-market flexibilization through adjustment of the legal framework. In these countries, it preceded or occurred concomitantly with the campaigns to attract foreign investment, so as to comfort MNC's decisions to invest in the country, but also to support job creation at times when these countries faced large pools of idle labor. Hence, this resulted in the higher bargaining power prescribed by law being followed by a rather declining labor share.

Table 4 – Results of the full model with qualitative bargaining indicators, all countries

| | Dependent variable: Log labor share | | | | | | | | | | | |
|---|---|---|---|---|---|---|---|---|---|---|---|---|
| | Fixed effects estimator | | | | | | IV/2SLS estimator | | | | | |
| | (1) | (2) | (3) | (4) | (5) | (6) | (7) | (8) | (9) | (10) | (11) | (12) |
| Capital-output ratio (log) | -0.176*** | -0.277*** | -0.302*** | -0.301*** | -0.307*** | -0.312*** | -0.186 | -0.295*** | -0.422*** | -0.409*** | -0.635*** | -0.464*** |
| | (0.044) | (0.054) | (0.055) | (0.055) | (0.062) | (0.055) | (0.146) | (0.065) | (0.061) | (0.083) | (0.157) | (0.092) |
| TFP (log) | -0.577*** | -0.605*** | -0.649*** | -0.655*** | -0.659*** | -0.663*** | -0.406 | -0.824*** | -0.843*** | -0.939*** | -0.988*** | -0.991*** |
| | (0.063) | (0.112) | (0.114) | (0.116) | (0.120) | (0.114) | (0.261) | (0.189) | (0.123) | (0.150) | (0.173) | (0.109) |
| Employment growth (log) | -1.458*** | -1.171** | -1.197** | -1.208** | -1.197** | -1.113** | -13.86** | -6.419 | -0.423 | 0.427 | -2.927 | -3.656 |
| | (0.546) | (0.534) | (0.541) | (0.528) | (0.532) | (0.541) | (5.930) | (6.569) | (4.026) | (6.380) | (5.450) | (4.045) |
| Basic requirements in collective bargaining | | 0.108*** | | | | | | 0.234 | | | | |
| | | (0.037) | | | | | | (0.181) | | | | |
| Basic characteristics of collective agreements | | | 0.0509** | | | | | | 0.366*** | | | |
| | | | (0.025) | | | | | | (0.089) | | | |
| Non-signatory parties and exemptions | | | | 0.0178 | | | | | | 0.689*** | | |
| | | | | (0.039) | | | | | | (0.254) | | |
| Institutional/legal characteristics of tripartite dialogue | | | | | 0.0226 | | | | | | 0.598*** | |
| | | | | | (0.055) | | | | | | (0.212) | |
| Members/representatives in tripartite dialogue | | | | | | 0.0667*** | | | | | | 0.279*** |
| | | | | | | (0.024) | | | | | | (0.086) |
| Constant | -1.805*** | -2.377*** | -2.339*** | -2.230*** | -2.298*** | -2.353*** | -2.297*** | -2.820*** | -3.324*** | -2.967*** | -5.007*** | -3.164*** |
| | (0.129) | (0.182) | (0.211) | (0.205) | (0.323) | (0.201) | (0.419) | (0.250) | (0.223) | (0.262) | (0.930) | (0.279) |
| Observations | 365 | 174 | 174 | 174 | 174 | 174 | 207 | 130 | 130 | 130 | 130 | 130 |
| R-squared | 0.437 | 0.409 | 0.385 | 0.374 | 0.374 | 0.393 | -2.024 | -0.194 | 0.274 | -0.44 | -0.175 | 0.187 |
| Hansen test (p-value) | | | | | | | 0.256 | 0.020 | 0.858 | 0.917 | 0.183 | 0.284 |

Source: Authors' estimates. The trade union density rate from the most recent past period and the log number of days since C154 has been ratified (0 if not ratified yet) used as exogenous instruments. Standard errors are provided in parentheses and are robust to arbitrary heteroskedasticity.

Table 5 – Results of the full model with qualitative bargaining indicators, transition countries

| | Dependent variable: Log labor share | | | | | | | | | | | |
|---|---|---|---|---|---|---|---|---|---|---|---|---|
| | Fixed effects estimator | | | | | | | IV/2SLS estimator | | | | |
| | (1) | (2) | (3) | (4) | (5) | (6) | (7) | (8) | (9) | (10) | (11) | (12) |
| Capital-output ratio (log) | -0.175*** | -0.278*** | -0.311*** | -0.309*** | -0.304*** | -0.325*** | -0.269*** | -0.254*** | -0.369*** | -0.344*** | -0.589*** | -0.444*** |
| | (0.044) | (0.054) | (0.046) | (0.043) | (0.057) | (0.056) | (0.086) | (0.076) | (0.055) | (0.058) | (0.092) | (0.050) |
| TFP (log) | -0.583*** | -0.604*** | -0.699*** | -0.734*** | -0.652*** | -0.683*** | -0.657*** | -0.917*** | -0.911*** | -0.977*** | -0.846*** | -0.884*** |
| | (0.066) | (0.114) | (0.097) | (0.099) | (0.115) | (0.112) | (0.127) | (0.204) | (0.122) | (0.129) | (0.194) | (0.126) |
| Employment growth (log) | -1.503*** | -1.214** | -0.648** | -0.387 | -0.946* | -0.899** | -8.364*** | -8.046 | -3.342 | -3.716 | 1.127 | -0.806 |
| | (0.579) | (0.510) | (0.283) | (0.301) | (0.517) | (0.359) | (2.160) | (8.864) | (4.589) | (4.893) | (6.337) | (3.697) |
| Transition economies (dummy) | -0.0674 | 1.035* | 1.521*** | 0.755*** | 1.142*** | 0.777*** | 0.234* | -6.757 | 2.927* | 0.874* | 1.55 | -0.0361 |
| | (0.081) | (0.555) | (0.228) | (0.127) | (0.429) | (0.145) | (0.135) | (6.255) | (1.657) | (0.480) | (1.735) | (0.972) |
| Basic requirements in collective bargaining | | 0.139* | | | | | | 0.528** | | | | |
| | | (0.072) | | | | | | (0.212) | | | | |
| *Transition economies* | | -0.325* | | | | | | 1.89 | | | | |
| | | (0.180) | | | | | | (1.998) | | | | |
| Basic characteristics of collective agreements | | | 0.135*** | | | | | | 0.235*** | | | |
| | | | (0.034) | | | | | | (0.050) | | | |
| *Transition economies* | | | -0.478*** | | | | | | -1.084* | | | |
| | | | (0.070) | | | | | | (0.614) | | | |
| Non-signatory parties and exemptions | | | | 0.208*** | | | | | | 0.356*** | | |
| | | | | (0.047) | | | | | | (0.078) | | |
| *Transition economies* | | | | -0.573*** | | | | | | -0.769** | | |
| | | | | (0.092) | | | | | | (0.371) | | |
| Institutional/legal char. of tripartite dialogue | | | | | 0.0592 | | | | | | 0.534*** | |
| | | | | | (0.059) | | | | | | (0.137) | |
| *Transition economies* | | | | | -0.321** | | | | | | -0.373 | |
| | | | | | (0.142) | | | | | | (0.576) | |
| Members/representatives in tripartite dialogue | | | | | | 0.131*** | | | | | | 0.254*** |
| | | | | | | (0.033) | | | | | | (0.046) |
| *Transition economies* | | | | | | -0.321*** | | | | | | 0.204 |
| | | | | | | (0.055) | | | | | | (0.533) |
| Constant | -1.778*** | -2.434*** | -2.580*** | -2.453*** | -2.465*** | -2.511*** | -2.479*** | -3.194*** | -3.061*** | -2.781*** | -4.655*** | -3.047*** |
| | (0.140) | (0.206) | (0.186) | (0.156) | (0.306) | (0.209) | (0.279) | (0.320) | (0.187) | (0.178) | (0.548) | (0.191) |
| Observations | 365 | 174 | 174 | 174 | 174 | 174 | 207 | 130 | 130 | 130 | 130 | 130 |
| R-squared | 0.439 | 0.419 | 0.527 | 0.529 | 0.419 | 0.474 | 0.257 | 0.455 | 0.523 | 0.465 | 0.26 | 0.522 |
| Hansen test (p-value) | | | | | | | . | 0.354 | 0.201 | 0.285 | 0.553 | 0.909 |

*Source*: Authors' estimates. The trade union density rate from the most recent past period and the log number of days since C154 has been ratified (0 if not ratified yet) used as exogenous instruments. Standard errors are provided in parentheses and are robust to arbitrary heteroskedasticity.

## 6. Conclusions

This paper aimed to shed some light on the way in which workers' bargaining power is relevant for the labor share, with strong focus on transition economies. Our theoretical foundation is the share-capital model borrowed from Bantolila and Saint Paul (2003), which prescribes three sources of variation in the labor share: shocks whose effect on the labor share is entirely mediated by the capital-output ratio and implies moves along the curve; the capital-augmenting technical progress which shifts the share-capital curve; and movements off the share-capital schedule, namely changes in labor adjustment costs and in workers' bargaining power, of which we are mainly interested in the latter.

We faced multiple constraints from the side of available data, because industry-level variables hold multiple time gaps, trade-union and collective-bargaining-coverage variables are not available for many developing countries, while bargaining enabling environment is available only in present times. Yet, we attempted to make use of the combined information to the extent possible. First, we used time-series information to estimate the basic share-capital schedule for the entire sample and the sample of transition economies, to have the capital-labor elasticity and the effect of the capital-augmenting technical progress as yardsticks against we verify results' robustness when estimates are later derived from a cross-sectional specification. Then, we estimate the full model, hence also involving the variables of interest – workers' bargaining power captured through quantitative and qualitative indicators, to capture the moves off the share-capital schedule.

Instrumental variable methods are used to estimate the empirical model, whereby former values of trade unionization rates and the time elapsed since the country adopted the ILO Collective Bargaining Convention, 1981 (No. 154) have been used as instruments.

The paper is the first to make use of own-constructed indices capturing the workers' bargaining power in qualitative sense, through coding textual information from the ILO IRLex database.

Results robustly reveal a capital-labor elasticity above one, and within slightly wider interval for transition economies. TFP is found with a strictly capital-augmenting role. Higher labor adjustment costs reduce labor share, but the result is nor stable.

Bargaining power of workers has a significant role to play for the labor share. In general, results across specifications suggest that the stronger the bargaining, the higher the labor share. However, the conclusion upturns in transition economies: generally, higher workers' bargaining power results in lower labor share. One plausible explanation rests with the notion of delayed response of the labor share to wage pushes attained through workers' concerted effort. It likely reconciles reality whereby transition economies experienced influx of foreign capital after the dissolution of the former socialist system, which did not confront the workers' power rise per se, hence not directly impacting the bargaining power reflected in unionization rates or on paper, but operated through its underlying channels, like the rapid automation MNCs introduced in their settlements in the host countries, and the change in market structure towards larger and more powerful firms, accompanied by a concerted effort of host governments to flexibilize labor markets and comfort MNCs' decisions to invest, ultimately overturning the influence of bargaining power onto labor share or adjourning its potentially positive response.

Annex

### Table A1 – Descriptive statistics for the UNIDO-based indicators

| Variable | Obs | Mean | Std. Dev. | Min | Max |
|---|---|---|---|---|---|
| Wages (million current LCU) | 41,360 | 1,500 | 5,760 | 401 | 94,700 |
| Value added (million current LCU) | 9,829 | 2,400 | 8,580 | -864 | 132,000 |
| Output (million current LCU) | 8,983 | 6,000 | 18,900 | 2 | 253,000 |
| Gross fixed capital formation (million current LCU) | 26,221 | 1,440 | 70,400 | -14,000 | 11,300,000 |
| Employment (number) | 38,326 | 101,207 | 462,236 | 1 | 10,200,000 |

### Table A2 – Descriptive statistics for the ILO-based indicators

| Variable | Obs | Mean | Std. Dev. | Min | Max |
|---|---|---|---|---|---|
| Trade union density rate (%) | 18,281 | 23 | 17 | 0.2 | 92 |
| Collective bargaining coverage rate (%) | 13,931 | 38 | 31 | 0.4 | 99 |
| Number of strikes and lockouts | 9,980 | 138 | 313 | 0 | 2,127 |
| Days not worked due to strikes and lockouts (thousands) | 9,879 | 610 | 2,020 | 0 | 20,700 |
| Workers involved in strikes and lockouts, total (thousands) | 9,236 | 13,646 | 108,472 | 0 | 1,284,678 |
| Workers involved in strikes and lockouts, manufacturing (thousands) | 5,412 | 2,430 | 14,548 | 0 | 179,173 |

### Table A3 – Coding of the IRLex-derived qualitative indicators

| Indicator | | Coding rule |
|---|---|---|
| Representation requirements for trade unions to negotiate a collective agreement at national level | 0 | no data |
| | 1 | there are specific requirements |
| | 2 | some general provisions apply |
| | 3 | no requirements apply |
| Requirements for bargaining in good faith | 0 | no data |
| | 1 | no provisions for bargaining in good faith apply |
| | 2 | provisions for bargaining in good faith are nonobligatory |
| | 3 | provisions for bargaining in good faith are determined by law |
| | 4 | provisions for bargaining in good faith are specified and determined by law |
| Registration of collective agreements | 0 | no data |
| | 1 | subject to registration and revision by the respective ministry |
| | 2 | subject to registration |
| | 3 | not subject to registration |
| Initiating party | 0 | no data |
| | 1 | public body (ministry, government) |
| | 2 | both Ministry/gov't and social partners |
| | 3 | private body (social partners) |
| Duration of collective agreements | 0 | no data |
| | 1 | Not exceeding 1 year |
| | 2 | Not exceeding 2 years |
| | 3 | Not exceeding 3 years |
| | 4 | Not exceeding 4 years |
| | 5 | Not exceeding 5 years |
| | 6 | Decided by parties |

| | | | |
|---|---|---|---|
| Amendment of collective agreements | 0 | no data | |
| | 1 | no option for amendment | |
| | 2 | written request to public body; no change unless expressly stipulated in the agreement | |
| | 3 | written request to parties; termination and submit a draft | |
| | 4 | as stipulated in the agreement | |
| | 5 | consensual decision | |
| Termination of collective agreements | 0 | no data | |
| | 1 | not possible to terminate / denunciate / the court / ministry may terminate if deemed necessary | |
| | 2 | in writing to the parties and/or labor inspectorate/ministry | |
| | 3 | in writing at the behest of any party / any party may terminate by giving prior notice | |
| | 4 | consensual decision | |
| | 5 | as stipulated in the agreement | |
| Extension to non-statutory parties (automatic/non-automatic) | 0 | no data | |
| | 1 | if not automatic (requires involvement of concerned parties, public body or a ministerial decision) | |
| | 2 | both apply | |
| | 3 | automatic extension | |
| Extension to non-statutory parties (resting decision) | 0 | no data | |
| | 1 | upon ministerial decision | |
| | 2 | extension upon request to a public body / requires approval of a minister | |
| | 3 | extension upon written request to parties / requires mutual consent | |
| | 4 | as stipulated in the agreement / if agreement concerns a certain number of employees in the profession / other general provisions | |
| Mechanisms for exemption from binding effects of an extended collective agreement | 0 | no data | |
| | 1 | no option for an exception | |
| | 2 | submitting a written request to a public body / By approval of the minister | |
| | 3 | some written form / signing an annex to the existing agreement | |
| | 4 | as stipulated in the agreement / meet certain requirements | |
| | 5 | very low level of rigidity: consensual decision | |
| Number of institutions acting as an umbrella for tripartite dialogue | 0 | no data | |
| | 1 | no institutions | |
| | 2 | one institution | |
| | 3 | two institutions | |
| | 4 | three institutions | |
| Legal status of the institution | 0 | no data | |
| | 1 | the institution is a body under a purview of a ministry or a public agency (public body) | |
| | 2 | the institution is an independent body | |
| Founding instrument | 0 | no data | |
| | 1 | an agreement | |
| | 2 | a by-law | |
| | 3 | a law | |
| | 4 | a decree | |
| | 5 | a constitution | |
| Legal effect of the institutions | 0 | no data | |
| | 1 | non-binding | |

| | 2 | binding |
|---|---|---|
| Number of members in the founding body | 0 | no data |
| | 1 | 1 member but no more than 60 |
| | 2 | 61 members but no more than 120 |
| | 3 | 121 members but no more than 180 |
| | 4 | 181 members but no more than 240 |
| | 5 | 241 members but no more than 300 |
| | 6 | 301 members but no more than 360 |
| Representatives other than workers, employers and government included in the institution | 0 | no data |
| | 1 | no other members |
| | 2 | can take part by invitation |
| | 3 | active members branched by divisions only |
| | 4 | active members branched by divisions and number of members per division |

Table A4 – Descriptive statistics for the IRLex-based indicators

| Variable | Obs | Mean | Std. Dev. | Min | Max |
|---|---|---|---|---|---|
| Indicators | | | | | |
| Representation requirements for trade unions to negotiate a collective agreement at national level | 70 | 0.80 | 1.06 | 0 | 3 |
| Requirements for bargaining in good faith | 70 | 1.97 | 1.51 | 0 | 4 |
| Registration of collective agreements | 70 | 1.46 | 0.83 | 0 | 3 |
| Initiating party | 70 | 1.51 | 1.35 | 0 | 3 |
| Duration of collective agreements | 70 | 2.74 | 2.41 | 0 | 6 |
| Amendment of collective agreements | 70 | 2.56 | 1.98 | 0 | 5 |
| Termination of collective agreements | 70 | 1.24 | 1.64 | 0 | 5 |
| Extension to non-statutory parties (automatic/non-automatic) | 70 | 1.01 | 1.00 | 0 | 3 |
| Extension to non-statutory parties (resting decision) | 70 | 1.24 | 1.56 | 0 | 4 |
| Mechanisms for exemption from binding effects of an extended collective agreement | 70 | 0.44 | 1.12 | 0 | 5 |
| Number of institutions acting as an umbrella for tripartite dialogue | 70 | 2.36 | 0.68 | 0 | 4 |
| Legal status of the institution | 70 | 0.97 | 0.74 | 0 | 2 |
| Founding instrument | 70 | 2.79 | 1.37 | 0 | 5 |
| Legal effect of the institutions | 70 | 0.89 | 0.50 | 0 | 2 |
| Number of members in the founding body | 70 | 1.01 | 0.77 | 0 | 4 |
| Representatives other than workers, employers and government included in the institution | 70 | 1.80 | 1.42 | 0 | 4 |
| Indices | | | | | |
| Basic requirements in collective bargaining | 70 | 2.49 | 1.17 | 0 | 4.83 |
| Basic characteristics of collective agreements | 70 | 2.58 | 1.36 | 0 | 5.10 |
| Non-signatory parties and exemptions | 70 | 1.47 | 1.20 | 0 | 4.67 |
| Institutional/legal characteristics of tripartite dialogue | 70 | 3.11 | 1.14 | 0 | 4.88 |
| Members/representatives in tripartite dialogue | 70 | 1.86 | 1.29 | 0 | 5.00 |